\documentclass[
journal=nalefd, 
manuscript=article]{achemso}

\usepackage[version=3]{mhchem} 
\usepackage{nicefrac}
\usepackage{xfrac}

\usepackage{amsmath} 

\usepackage[utf8]{inputenc}
\usepackage[english]{babel}

\usepackage[usenames, dvipsnames]{color}
\usepackage{titlesec}

\titleformat{\section}
{\normalfont\fontsize{12}{15}\bfseries}{\thesection}{1em}{}

\usepackage{gensymb}

\author{Abdullah Alharbi}
\author{Darren Armstrong}
\author{Somayah Alharbi}
\affiliation[NYU]
{Department of Electrical and Computer Engineering, New York University, Brooklyn, NY }

\author{Davood Shahrjerdi}
\email{davood@nyu.edu}
\affiliation[NYU]
{Department of Electrical and Computer Engineering, New York University, Brooklyn, NY }

\title [\texttt{achemso} primitives]
{Physical cryptographic primitives by chemical vapor deposition of layered \ce{MoS2}
}


\begin{document}
	
	\begin{abstract}
		Development of physical cryptographic primitives for generating strong security keys is central to combating security threats such as counterfeiting and unauthorized access to electronic devices.  We introduce a new class of physical cryptographic primitives from layered molybdenum disulfide (\ce{MoS2}) which leverages the unique properties of this material system.  Using chemical vapor deposition, we synthesize a \ce{MoS2} monolayer film covered with speckles of multilayer islands, where the growth process is engineered for an optimal speckle density. The physical cryptographic primitive is an array of 2048 pixels fabricated from this film.  Using the Clark-Evans test, we confirm that the distribution of islands on the film exhibits complete spatial randomness, making this cryptographic primitive ideal for security applications. A unique optical response is generated by applying an optical stimulus to the structure. The basis for this unique response is the dependence of the photoemission on the number of \ce{MoS2} layers which by design is random throughout the film. The optical response is used to generate cryptographic keys. Standard security tests confirm the uniqueness, reliability, and uniformity of these keys.  This study reveals a new opportunity for generating strong and versatile nano-engineered security primitives from layered transition metal dichalcogenides.
	\end{abstract}

	\maketitle
	
	\vspace{1cm}
	
	Modern society demands information security~\cite{Hardware2017}. Globalization of supply chains has undermined trust in electronic devices, which were once manufactured entirely by a single \textit{trusted} factory. Further, the ubiquity of today's advanced manufacturing poses additional challenges, because such resources are now more accessible to adversaries for developing sophisticated security attacks. A vast number of such attacks are physical~\cite{rostami2013hardware, Karri2014} and range from counterfeiting to various forms of unauthorized access such as reverse engineering and side-channel attacks. Authentication of electronic devices through unique security keys has become the first line of defense. A security key must be easy to generate and yet impossible to replicate. An increasingly popular method of generating security keys is based on \textit{physical} cryptographic primitives which leverage the inherent randomness of a structure or a physical process (\textit{e.g.} manufacturing variability or materials disorders)~\cite{pappu2002physical,pham2007polymer,huang2010unbreakable, han2012,blumenthal2012patterned,kim2014anti, silicon2002}. Applying a challenge (such as an electrical or an optical stimulus) to the cryptographic primitive produces a unique response (a security key). Hence, this concept produces cryptographic keys that are unique for each electronic device. 
	
	\begin{figure}[t]
		\centering
		\includegraphics[width=0.9\textwidth]{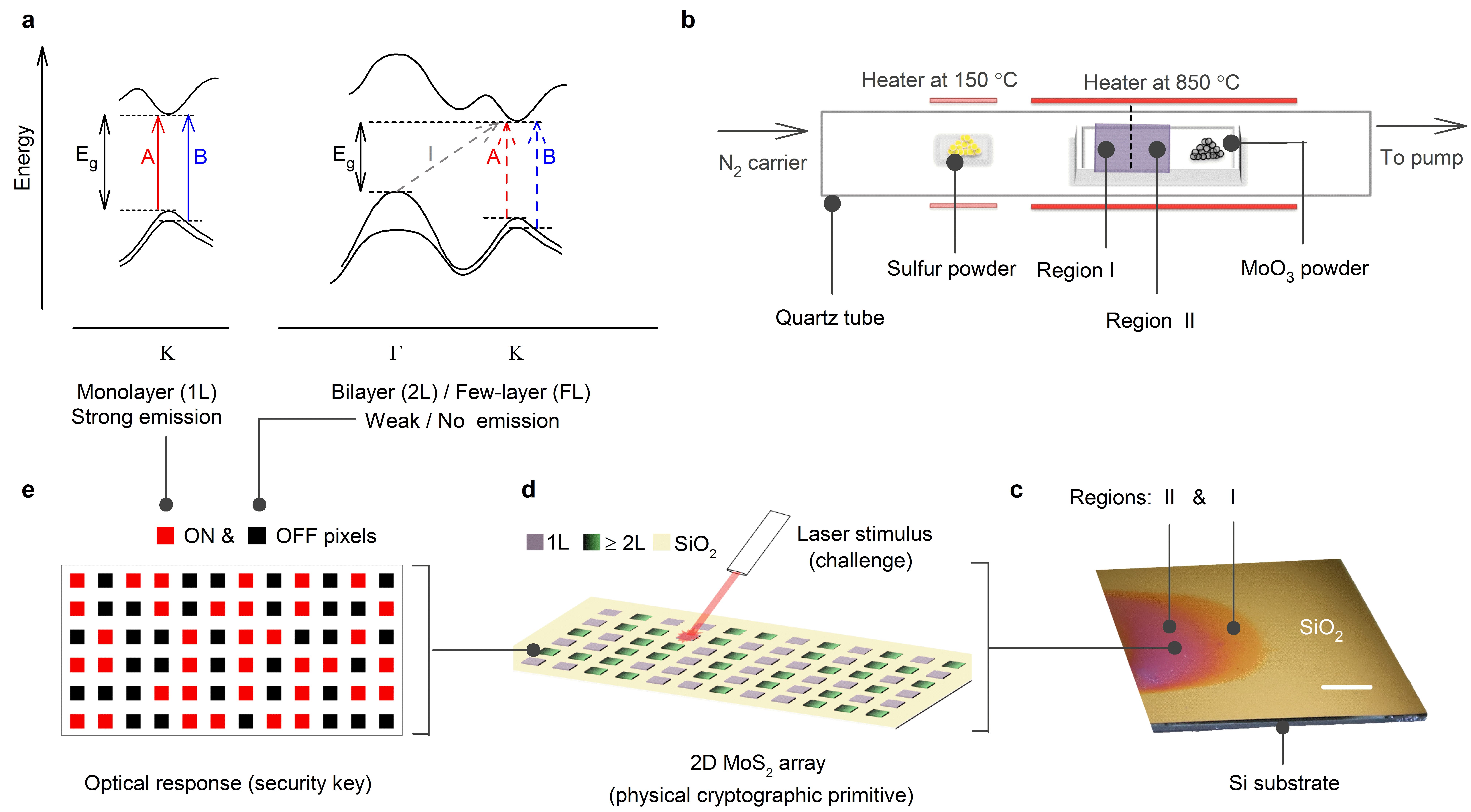}
		\caption{\textbf{\ce{MoS2}-based physical cryptographic primitives:}~(a) Schematic illustration of the energy band structure of monolayer and multilayer \ce{MoS2}, indicating strong dependence of excitonic emission on the number of \ce{MoS2} layers. (b) CVD using solid-phase precursors was used for synthesis of large-area \ce{MoS2} films. (c) Photo of a CVD \ce{MoS2} sample indicating two distinct regions of growth. Scale bar is 5~mm. (d) An array of $32\times 64$ pixels was then formed in region II. (e) Stimulating the physical \ce{MoS2} primitive with a laser light produces a unique optical response with randomly distributed ON and OFF pixels.}
		\label{fig:concept}  
	\end{figure}
	
	At present, silicon-based security primitives are widely used due to their compatibility with the complementary metal-oxide-semiconductor (CMOS) technology \cite{silicon2002}. However, the device variability begins to diminish as a given CMOS technology matures, thereby reducing the effectiveness of these silicon cryptographic primitives. In fact, mature CMOS technologies are increasingly used in many applications ranging from automotive industry to smart gadgets for the so-called Internet-of-Things, thereby leaving them vulnerable to the growing security threats. It is also infeasible for silicon-based primitives to secure many emerging technologies simply due to practical considerations such as material compatibility, robust operation, and cost. Flexible electronics and self-powered sensor nodes are prime examples for such emerging technologies. Hence, there is a need for developing new non-silicon cryptographic primitives that are strong, robust, and versatile.
	
	Nanomaterials offer distinct physical properties which are often nonexistent in silicon. In the past two decades, the continual discovery of new nanomaterials, from carbon nanotubes to two-dimensional (2D) transition metal dichalcogenides (TMDs), has formed the basis for creating the next-generation electronics that are high-speed and low-power~\cite{high2010,lowpower2013, bhimanapati2015recent}. However, the variability of nanomaterials remains a practical barrier for making these devices on a large scale~\cite{franklin2012variability,park2012high}. On the other hand, this randomness provides an opportunity for making physically unclonable security primitives. This concept is experimentally demonstrated by the recent work of Hu \textit{et al.} which implements a new electrical cryptographic primitive using the randomness of a carbon nanotube assembly process~\cite{CNT2016}. Beyond carbon nanotubes, the prospects of layered TMDs for security applications are still unexplored.
	
	Here, we introduce a physical cryptographic primitive constructed from layered molybdenum disulfide (\ce{MoS2}). We used \ce{MoS2} as the model system since it has been heavily studied in the family of 2D layered TMDs~\cite{atomically2010, emerging2010}. Figure~\ref{fig:concept} illustrates the proposed concept. Our approach combines two fundamental phenomena for producing physically unclonable \ce{MoS2}-based primitives. The first phenomenon is the inherent difference in excitonic emission strengths of a monolayer and a multilayer \ce{MoS2}, an attribute unique to most semiconducting TMDs as shown in figure~\ref{fig:concept}a. The second phenomenon relates to complete spatial randomness of an ideal island growth, that is inherent to thin film growth techniques. A large-area \ce{MoS2} film is produced using a chemical vapor deposition (CVD) process in a layer-plus-island growth mode (figure~\ref{fig:concept}b). Figure~\ref{fig:concept}c shows the photo of a CVD \ce{MoS2} film, illustrating two  distinct growth regions on the substrate. The region of interest (\textit{i.e.} region II) is composed of a continuous \ce{MoS2} monolayer (1L) with speckles of multilayer (bilayer 2L or few-layer FL) islands. We engineer the growth process to achieve an optimal island density in this region.  We confirmed the spatial randomness of the multilayer islands using the statistical test by Clark-Evans~\cite{clark1954distance}.  The physical cryptographic primitive itself is fabricated as a 2048-pixel array from the film in this region (figure~\ref{fig:concept}d).  Application of an optical (laser) stimulus to the security primitive results in random ON and OFF pixels (figure~\ref{fig:concept}e), owing to the spatial randomness of the multilayers and the different photoemission strengths of monolayer and multilayer pixels.  Using standard security tests, we confirm the randomness and stability of security keys generated from the proposed physical cryptographic primitive.       
	
	\section*{Layer-plus-island growth of CVD \ce{MoS2}} 
	
\begin{figure}[t]
		\centering
		\includegraphics{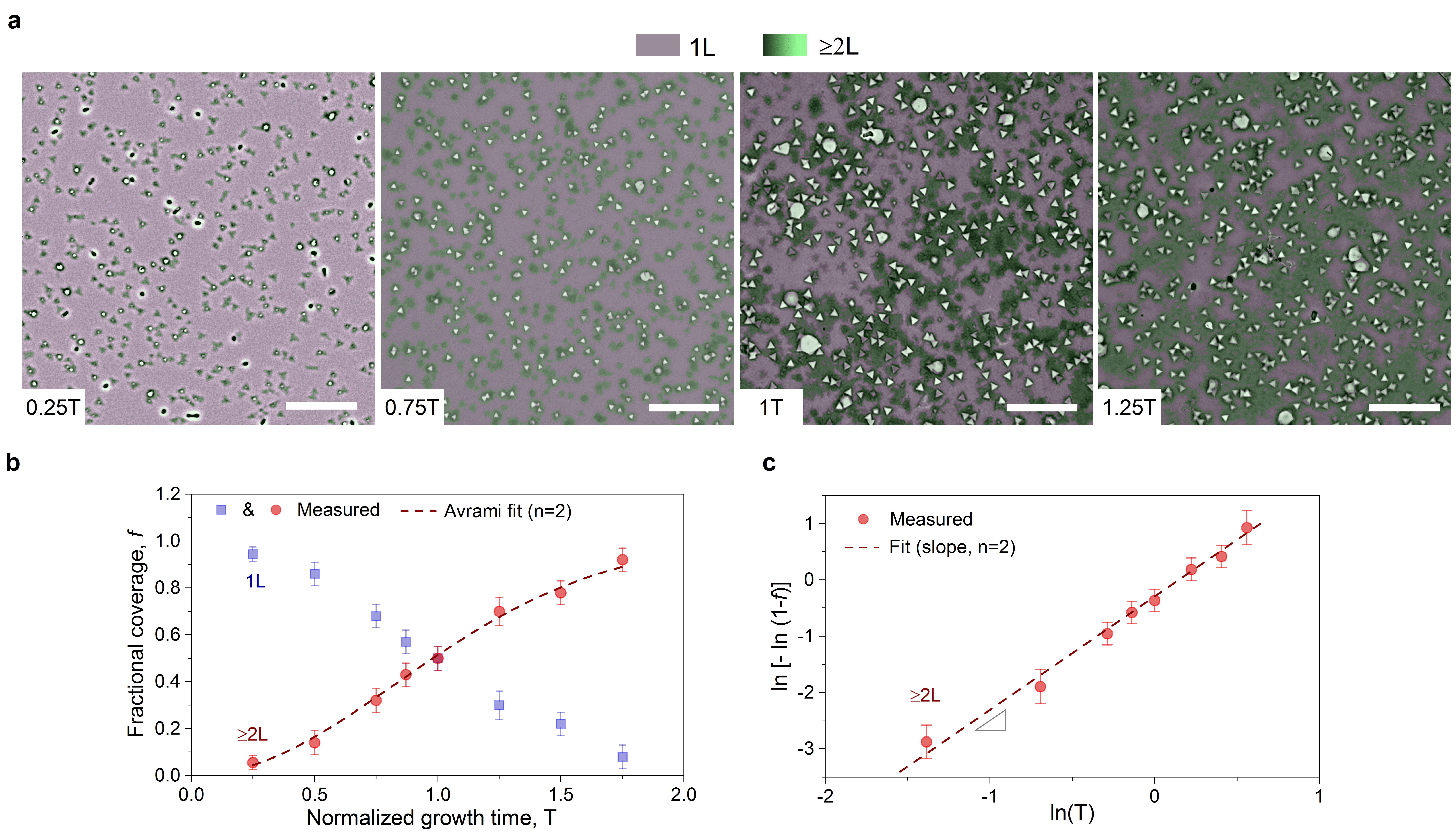}
		\caption{\textbf{Optimizing the growth randomness:}~ (a) Optical images illustrating the time evolution of the \ce{MoS2} growth in region II, indicating the layer-plus-island growth modes. (b) Time-dependent normalized surface coverage of monolayer (1L) and multilayer ($\ge$2L) films in region II. The Avrami equation provides a reasonable fit to the data, further confirming complete spatial randomness of island nucleation in region II. (c) The fit gives an Avrami exponent, $n$ of about 2, suggesting 2D disk-shaped growth governed by the surface diffusion. All scale bars are 30 $\mu m$}
		\label{fig:growth}
	\end{figure}

	Among the different methods for growing TMDs, CVD techniques have shown better thickness control on a large scale~\cite{van2013grains, yu2013controlled,kang2015high,alharbi2016electronic}. We grew large-area \ce{MoS2} films onto 285 nm \ce{SiO2} on $p^+$ silicon substrates by CVD from sulfur and \ce{MoO3} precursors~\cite{alharbi2017material}. Figure~\ref{fig:concept}b schematically illustrates the CVD reactor based on the solid-phase precursors. Two distinct growth regions are typically evident along the substrate (see figure~\ref{fig:concept}c), indicating that in this CVD process the growth depends on the distance of the substrate from the \ce{MoO3} powder. Region I is the farthest from the \ce{MoO3} powder in the reactor, where the optimal growth conditions yield a continuous monolayer. Region II, located closer to the \ce{MoO3} powder, is covered by a continuous monolayer film with randomly distributed multilayer islands. According to the surface science of thin film growth, the growth mode strongly depends on the deposition rate of the growth species and the substrate temperature~\cite{growth1997}. It is known that the growth mode will deviate from the layer-by-layer mode to the layer-plus-island mode once the deposition rate exceeds a critical value. This explains the presence of these two prominent growth modes along the substrate~\cite{kang2015high,growth1997}. Indeed, the layer-by-layer growth occurs in region I with low Mo vapor pressure and the growth follows a site-saturated growth kinetics (see Supplementary Information). Despite the spatial randomness of the nucleation sites, region I of our samples is sub-optimal for constructing a dense array of random binary code because of the the relatively sparse spatial distribution of the multilayer films grown mostly at the grain boundaries (spacing from 20-80 $\mu m$). In contrast, region II (closer to the \ce{MoO3} powder) is exposed to a higher concentration of \ce{Mo} vapor, resulting in the layer-plus-island growth mode and thus the random nucleation of multilayer islands on the monolayer film, as shown in figure~\ref{fig:growth}a. To produce the physical cryptographic primitive, we engineer the growth process in this region to achieve an optimal surface coverage of the multilayer islands. 
	
	\section*{Engineering and testing growth randomness} 
	
	Complete spatial randomness (CSR) is central to constructing strong cryptographic keys in our proposed concept. To test for CSR of the island growth in region II, we apply the statistical test by Clark and Evans~\cite{clark1954distance} on images from this region, taken at an early stage of the multilayer nucleation on the continuous monolayer film, \textit{e.g.} figure~\ref{fig:growth}a at $T=0.25$. If the island growth is CSR, then the distribution of nearest neighbor distances (\textit{i.e.} the distances between the islands and their nearest neighbor) has a mean $r_{CSR} = 1/(2\sqrt{\rho})$ and a variance $\sigma_{CSR}^2 = (4-\pi)/(4\pi\rho)$, where $\rho$ is the particle density per unit area.  Therefore, we can test for CSR by testing the null hypothesis that the mean of nearest neighbor distances is equal to $r_{CSR}$.  Using the two-tailed test for the population mean, we compute the standard $Z$-score given by:
	\begin{equation} \label{Z-test}
	Z = \frac{\langle r_s \rangle - \langle r_{CSR} \rangle}{\sqrt{\sigma_{CSR}^2/N}},
	\end{equation} 
	where $r_s$ is the sample mean of the nearest neighbor distances computed with $N$ particles. That is, 
	\begin{align}
	r_s = \sum_{i=1}^N \frac{r_i}{N},
	\end{align}
	where $r_i$ is the nearest neighbor distance of the $i$th island.  At a 0.05 significance level, the null hypothesis is to be rejected if $Z \leq -1.96$ or $Z \geq 1.96$.
	We calculated typical Z values of about 0.7--1.0 for our samples (see Supplementary Information). Hence, at the 0.05-level of significance, we cannot reject the null hypothesis that the mean nearest neighbor distance is $r_{CSR}$.  This suggests that the island growth in region II exhibits CSR, hence all the nucleations are independent and the probability of nucleation is the same everywhere on the surface.
	
	Considering CSR island growth in region II, the Avrami equation~\cite{avrami1939kinetics} can be used to draw insight into the time evolution of island growth. For a growth time $t$, the fractional surface coverage~$f$ by the multilayer islands is approximated by: 
	
	\begin{equation} \label{Avrami}
	f(t) = 1-e^{-kt^{n}}
	\end{equation} 
	
	\noindent where the Avrami exponent $n$ gives information about the kinetics of the island growth. To analyze the growth kinetics, we prepared several samples with varying growth times while keeping the other processing conditions identical including the sample dimensions and its position relative to \ce{MoO3}. We then imaged the samples to compute the surface coverage in region II, as shown in figure~\ref{fig:growth}a. Assuming time-invariant growth kinetics, this experiment provides a good approximation of the time evolution of the surface coverage~\cite{starink2001meaning}. From the optical images, we made two key observations: (i) nucleation is continuous evident from the concurrent presence of thin (mostly 2L) and thick islands in all different stages of the growth, and (ii) the growth is mostly 2D, \textit{i.e.} the lateral dimensions of islands grow faster than the thickness. Figure~\ref{fig:growth}b summarizes the time evolution of the normalized surface coverage $f$ for the monolayer film and the multilayer islands. In this plot, T is the normalized growth time, defined as $T=(t-t_0)/t_{0.5}$, where $t_{0}$ denotes the approximate growth time at which the surface is fully covered by a continuous monolayer and $t_{0.5}$ represents the time at which 50\% of the monolayer surface area is exposed and the rest is covered by the multilayer islands. In figure~\ref{fig:growth}c, we plotted $\ln [-\ln (1-f)]$ as a function of $\ln (T)$, where the slope of the fitted line gives the estimate for the Avrami exponent $n$. We found $n \approx 2$ for our growth experiments, suggesting a 2D disk-shaped growth governed by the surface diffusion. Equation~\ref{Avrami} provides a reasonable fit to the experimental data in figure~\ref{fig:growth}b, further confirming CSR of the growth in region II. Further, the inflection point of the fitted curve at $T \approx 0.75$ corresponds to the cross-over from the isolated island growth to the island overlap growth.
	
	After analyzing the growth kinetics, we adjusted the growth time to obtain \ce{MoS2} films with equal surface areas of exposed monolayer film and of the multilayer islands, \textit{i.e.} $f = 0.5$.  This was done to achieve the maximum randomness in the physical cryptographic primitive and in the security key responses.  
	
	\section*{Cryptographic key generation} 
	
\begin{figure}[t]
		\includegraphics{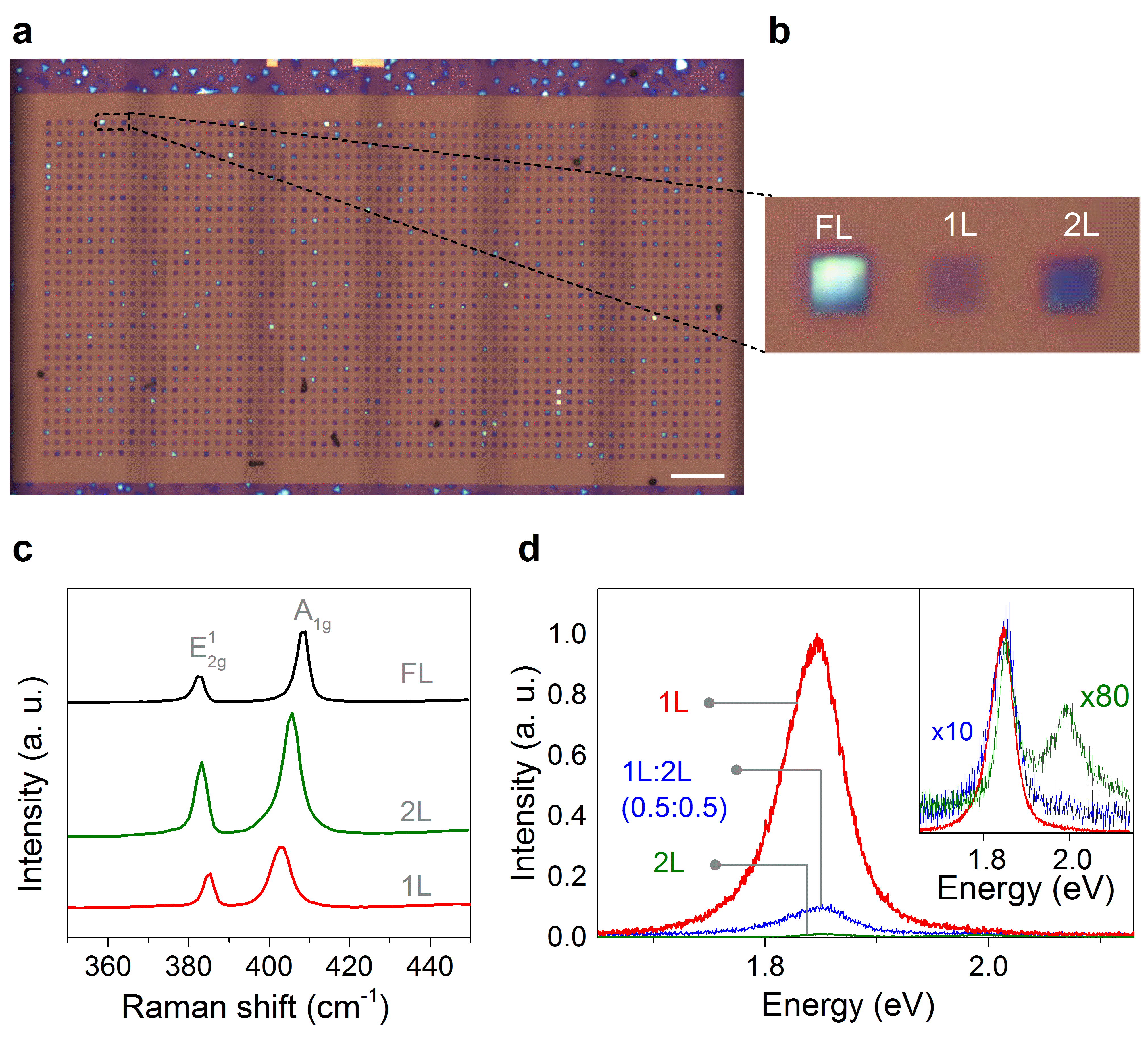}
		\caption{\textbf{\ce{MoS2} physical cryptographic primitive:}~(a) Optical image of a 2D array with 2048 \ce{MoS2} pixels. (b) Zoomed-in optical image of three neighboring \ce{MoS2} pixels with different layer thickness. Corresponding (c) Raman spectra of the pixels with 1L, 2L, and FL thickness. (d) The PL characteristics of three pixels covered with 1L, 2L, and mixture of 1L:2L \ce{MoS2}. From the data, we set the ON/OFF classification threshold to 0.1.}
		\label{fig:fig3}
	\end{figure}

	To implement the \ce{MoS2} physical cryptographic primitives, we fabricated dense arrays consisting of 32$\times$64 pixels from the film in region II (see Methods). These arrays have a pixel size of 2 $\mu m\times$2 $\mu m$ and an equal pixel spacing of 2 $\mu m$. This pixel size was chosen because it is comparable with the dimensions of the state-of-the-art CMOS image sensors~\cite{ahn20147}. Figure~\ref{fig:fig3}a shows the optical image of a 2D \ce{MoS2} array with 2048 pixels, fabricated on a \ce{SiO2}/Si substrate. Due to the randomness of the nucleation, the content of each pixel is random. Specifically, a pixel might consist of a monolayer, a multilayer, or a mixture of the two. Figure~\ref{fig:fig3}b illustrates the zoomed-in view of three neighboring pixels. These pixels visually look different from one another, indicating their thickness difference. The Raman fingerprint of these pixels in figure~\ref{fig:fig3}c confirms the material type (which is \ce{MoS2} here) and the corresponding thickness, determined from the distance between the peak position of the in-plane ($E_{2g}^1$) and the out-of-plane ($A_{1g}$) phonon modes. 
	
\begin{figure}[t]
		\centering
		\includegraphics{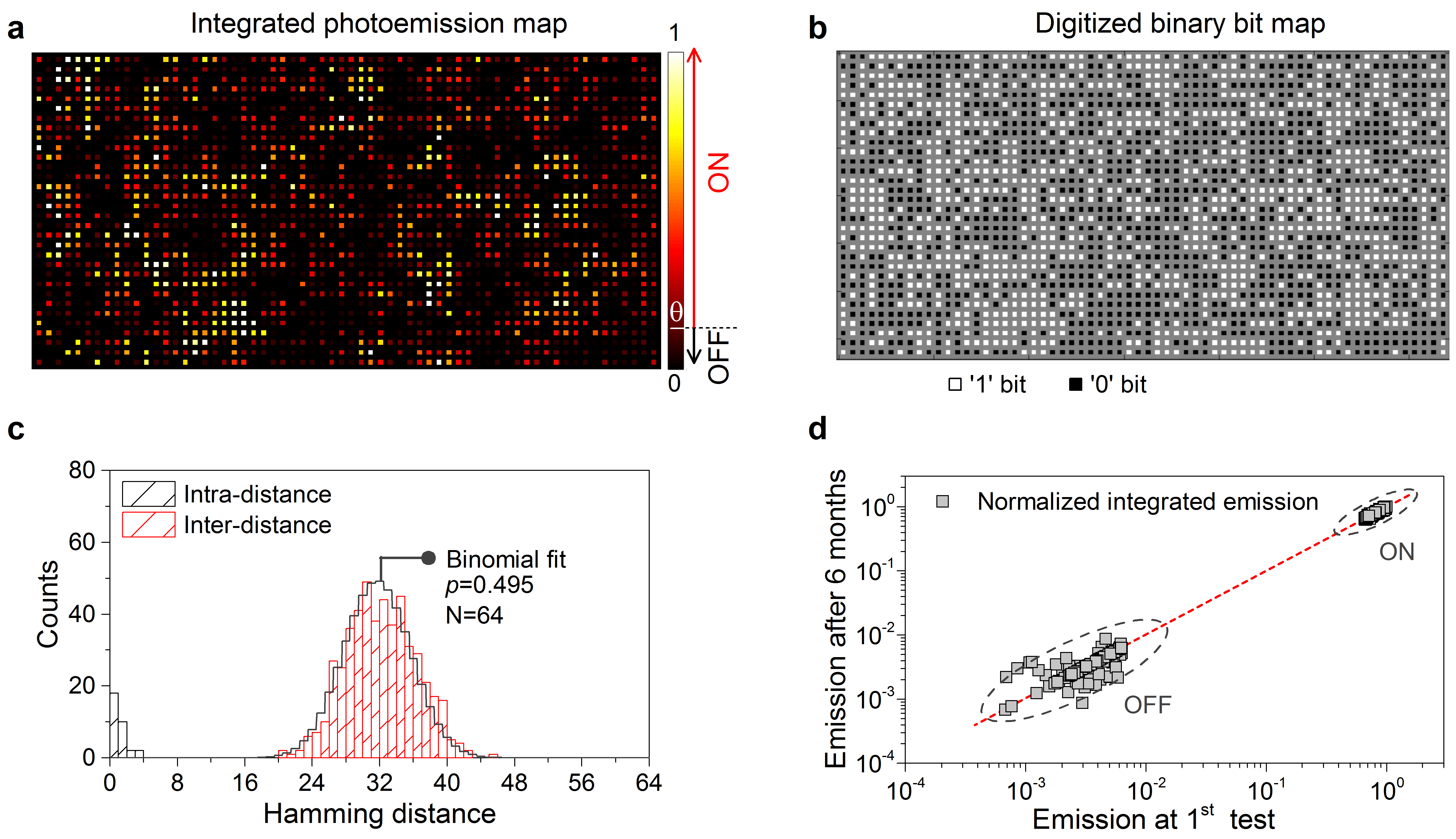}
		\caption{\textbf{Optical response and security metrics of the \ce{MoS2} primitive:}~(a) Stimulating the 2D \ce{MoS2} array with a laser light produces a optical response that is unique to this primitive. (b) The photoemission spatial map was converted to a 2D binary array by comparing each pixel with the ON/OFF threshold. (c) Standard security tests confirm uniqueness and repeatability of the security keys. (d) The \ce{MoS2} primitives are highly stable.}
		\label{fig:fig4}
	\end{figure}

	After fabrication, the physical security primitive was stimulated using a laser light to generate an optical response.  We expect the response to be unique to the cryptographic primitive given the random thickness distribution of the CVD \ce{MoS2} and the thickness dependence of the excitonic emissions in \ce{MoS2}.
	Specifically, pixels covered mostly by a monolayer film are expected to exhibit strong photon emission (ON pixels), while pixels mostly comprising of a multilayer film are OFF due to their weak emission properties. Figure~\ref{fig:fig3}d compares the typical photoluminescence (PL) spectra for three pixels comprising of: (i) full monolayer, (ii) full bilayer, and (iii) 50\% monolayer and 50\% bilayer. The corresponding optical images are shown in Supplementary Information. The PL spectra were normalized relative to that of the monolayer film. Two key observations are made from this plot. First, the photoemission of the full monolayer pixel is noticeably stronger than the pixel with full bilayer film. Second, the photoemission of the pixel with 50\% monolayer coverage is about 1/10th of the pixel with full monolayer.  Considering that the photoemission of a bilayer film is stronger than that of a film with three or more layers, the mixed monolayer-bilayer pixel represents the most ambiguous case for classifying a pixel as ON or OFF within an array. Therefore, for ON/OFF classification of the pixels, we set the threshold $\theta$ of the normalized photoemission to~0.1.
	We measured the corresponding PL spectrum of each pixel and then calculated the total area under the PL emission curve in the wavelength range of 580-770~nm (the integrated photoemission). Figure~\ref{fig:fig4}a is the spatial map of the normalized integrated photoemission for a 2D \ce{MoS2} array. We then converted the photoemission map to a 2D array of zero and one binary bits by comparing the normalized integrated emission of each pixel with the ON/OFF threshold $\theta$ of 0.1. The extracted 2D random binary code is shown in figure~\ref{fig:fig4}b. 
	
	Considering CSR of the island growth and the equal surface coverage by the monolayer and multilayer \ce{MoS2}, it is however expected that the distribution of the ON and OFF pixels shows no or weak dependency on the pixel size and the pixel spacing in the 2D \ce{MoS2} array. We confirmed this by fabricating multiple arrays with different pixel sizes and spacings, where the arrays demonstrate equal distribution of random ON and OFF pixels (see Supplementary Information). Hence, the strength of the cryptographic primitive is robust to the pixel choice and spacing choice.
	
	\section*{Analyzing security and stability metrics}
	
	We next analyzed the security metrics of the 2D binary array. Three important metrics are typically used to evaluate the strength of a cryptographic primitive~\cite{mukhopadhyay2014hardware}: \textit{uniqueness}, \textit{repeatability}, and \textit{uniformity}. 
	
	\textit{Uniqueness} is the ability of a key to be distinguished from other keys.  We use the average Hamming inter-distance to quantify uniqueness. The Hamming inter-distance between two keys is the minimum number of bit substitutions required to transform one key to another. The 32 rows of the 2D binary array are 64-bit security keys to be tested.  We compute the Hamming inter-distance of all 496 possible pairs of keys (see Supplementary Information).  Figure~\ref{fig:fig4}c shows the Hamming inter-distance distribution.  A binomial distribution with parameters $p=0.495$ and $N=64$ provides a good fit based on the Kolmogorov-Smirnov test. The inverse of the binomial distribution at cumulative probability 0.05 is 25.  This means that for two randomly generated 64-bit keys, there is a 95\% probability that the keys differ in at least 25 bits.  Hence, there is a 95\% chance that it will require at least 64 choose 25 (or $4 \times 10^{17}$) worst-case number of attempts to guess an unknown key from another known key. 
	
	A random key must also produce a consistent response to a given input challenge. The difference in response of a given binary key to the same challenge is quantified by the Hamming intra-distance, which represents the \textit{repeatability} of the random binary code. Therefore, the ideal intra-distance is zero. Figure~\ref{fig:fig4}c shows the results of the Hamming intra-distance, indicating high repeatability of the \ce{MoS2} security keys. The observed bit error rates are measurement artifacts and originate from the limited spatial accuracy of the automated sample stage when measuring the array.   
	
	To maximize the combination randomness of a binary array, each pixel should have an equal probability of being ON or OFF.  That is, there should be a \emph{uniformity} in the distribution of ON and OFF pixels in the array, with an ideal proportion of~0.5. Uniformity is quantified by the Hamming weight of the key, defined as the number of bit substitutions to convert the key to an array of all zeros.  We calculate the normalized Hamming weight on all 32 64-bit rows of the 2D binary array, and found the average to be 0.48 (see Supplementary Information).  The uniformity of the binary array arises from engineering the CVD process to achieve equal surface coverage by monolayer and multilayer. 
	
	Finally, the emission properties of the physical cryptographic primitives are unchanged after 6 months storage in ambient air, confirmed by measuring a random sample of 200 pixels, shown in figure~\ref{fig:fig4}d. Those pixels were either fully covered by a monolayer or a multilayer \ce{MoS2} film. The data indicates that our \ce{MoS2} physical cryptographic primitives are highly stable. 
	
	\section*{Conclusions}
	
	We introduced a physical cryptographic primitive based on layered molybdenum disulfide (\ce{MoS2}). Two fundamental properties underlie this security technology: (i) complete spatial randomness of multilayer island growth during chemical vapor deposition (CVD) of \ce{MoS2}, and (ii) strong thickness dependence of photoemission in \ce{MoS2}. These security primitives are easy to produce on a large scale using CVD and yet impossible to duplicate because of complete spatial randomness of the multilayer island growth. The findings of this study can be readily extended for the development of physically unclonable primitives based on other semiconducting transition metal dichalcogenides.

	\section*{Acknowledgement}
	
	This work was supported in part by NSF award $\#$1638598. This research used resources of the Center for Functional Nanomaterials, which is a U.S. DOE Office of Science Facility, at Brookhaven National Laboratory under Contract No. DE-SC0012704.The authors acknowledge C. Black and J. Uichanco for helpful discussions. 
	
	\section*{Methods}
	
	We performed CVD growth using \ce{MoO3} and sulfur solid precursors without requiring a growth promoter. The growth was performed using a custom-made setup at 850 \degree C with a nitrogen flow of 10~sccm. The optimal quantities of \ce{MoO3} and sulfur precursors are about 6~mg and 100~mg. In all the experiments, the films were grown in the presence of excess sulfur. The 2D \ce{MoS2} array was fabricated using an e-beam lithography step followed by patterning in an \ce{CF4}/\ce{O2} plasma. The 2D arrays were stimulated using a green laser for producing an optical response. 
	
	\pagebreak

	\bibliography{Optical_ref}
	
\end{document}